\documentclass[pra,preprint,showpacs]{revtex4}

\usepackage{epsfig}
\usepackage{graphicx}
\usepackage{bm}
\usepackage{amsmath}
\usepackage{dcolumn}

\newcommand{\rcm}{\mbox{cm$^{-1}$}}
\newcommand{\Sch}{Schr\"{o}dinger}

\begin{document}

\title{A regularized inverted perturbation approach method: potential energy curve
of the 4$^1\Sigma^{+}_{\mathrm u}$ state in Na$_2$}

\author{A. Grochola}
\author{P. Kowalczyk}
\thanks{Corresponding author}
\email {pfkowal@fuw.edu.pl}
\affiliation{Institute of Experimental
Physics, Warsaw University, ul. Ho\.za 69, 00-681~Warsaw, Poland}
\author{W. Jastrzebski}
\affiliation{Institute of Physics, Polish Academy of Sciences,
Al.Lotnik\'{o}w 32/46, 02-668~Warsaw, Poland}
\author{A. Pashov}
\affiliation{Institute for Scientific Research in
Telecommunications, ul. Hajdushka poliana 8, 1612 Sofia, Bulgaria}

\date{\today}

\begin{abstract}
We describe a modification of the inverted perturbation approach
method allowing to construct physically sensible potential energy
curves for electronic states of diatomic molecules even when some
parts of the potential are not adequately characterized by the
experimental data. The method is based on a simple regularization
procedure, imposing an additional constraint on the constructed
potential curve. In the present work it is applied to the double
minimum 4$^1\Sigma^{+}_{\mathrm u}$ state of Na$_2$, observed
experimentally by polarization labeling spectroscopy technique.
\end{abstract}


 \maketitle

\section{Introduction}

Perhaps the most challenging among the electronic states of
diatomic molecules are those characterized by exotic shapes of
the potential energy curves, which may exhibit barriers towards
dissociation, unusual bends or shelves as well as multiple
minima. Most of these features reflect interactions between the
initially regular diabatic molecular states which, if belonging
to the same irreducible representation of the symmetry group of
the molecule, in the adiabatic approach cannot cross each other.
Except of satisfying a natural, fundamental interest,
investigation of such states can bring an immediate, twofold
profit. First, the complicated shape of the exotic curves is very
sensitive to relative positions of the original diabatic
potentials. Therefore experimentally determined curves provide
particularly rigorous tests of the quality of theoretical
calculations. Second, it happens frequently that exotic potential
wells are unusually wide. Such a well can be employed then as an
intermediate state to access the long range region of other
states situated below or above it.

The unusual electronic states require special methods of analysis
and reduction of the experimental data to potential energy
curves. In the simplest instance of regular Morse-like
potentials, the standard Rydberg-Klein-Rees (RKR) algorithm
\cite{A}, based on the Bohr-Sommerfeld quantization of the phase
integral for the vibrational motion, is used. This algorithm can
also be generalized to treat some ``irregular'' curves
\cite{Stwalley,36nak}, but evidently fails in the case of double
minimum potentials. As an alternative the variational inverted
perturbation approach (IPA) has been developed \cite{Kozman},
providing fully quantum mechanical determination of potential
energy curves directly from the experimental data. In this
procedure, one starts with some estimated approximate potential
curve and iteratively seeks corrections to it until the quantum
mechanical eigenenergies calculated from the improved curve agree
with the experimental term values in the least squares
approximation (LSA) sense. Several routines have been proposed
for realization of this idea, differing mainly in mathematical
representation of the potential
\cite{Vidal,weCPC,LeRoy,L1,L2,Tiemann}. In most of them
analytical expressions of various forms
\cite{Vidal,LeRoy,L1,L2,Tiemann} allow to accurately describe
potential energy curves with few fitting parameters, but up to
now they have been designed to deal only with ``regular''
potentials with a Morse-like shape.

The pointwise representation of the potential, proposed within
the framework of the IPA method by Pashov \textit{et al.}
\cite{weCPC}, is more general. As an essentially model-free
approach, it imposes no limitation on the shape of the fitted
curve and therefore can be applied to both "regular" and
"irregular" potential energy curves
\cite{K2dm,Li2F,KLiX,NaLi,NaK4S,KLiD}. However, the pointwise
approach has also disadvantages following from its model-free
nature. The method works well for the parts of the potentials
covered by abundant experimental data (i.e. energies of
rovibrational levels), but when the experimentally determined
energy levels become sparse, the inversion problem starts to be
highly ill-conditioned and the procedure may become unstable,
producing irregularities in a form of unphysical wiggles on the
constructed potential curve. In the original algorithm
\cite{weCPC} the problem was partially solved by using the
singular value decomposition (SVD) technique \cite{SVD} instead
of the standard LSA for fitting the molecular potential.
Additional smoothing of the potential resulted from using a cubic
spline function for interpolation between the points defining the
potential energy curve; increasing the grid spacing could reduce
the undesirable flexibility of the constructed curve.
Nevertheless in some cases fitting a smooth potential to few
experimental data turned out to be nontrivial and even tricky.

We have recently encountered such a problem when investigating the
$4^1\Sigma^+_{\mathrm u}$ state in Na$_2$ molecule. To the best
of our knowledge this state has escaped experimental
characterization up to now. The $^1\Sigma^+$ states in alkali
dimers are expected to show an intricate system of avoided
crossings between different Rydberg and valence states at short
internuclear distances and also ion-pair states (Na$^+$Na$^-$) at
long internuclear distances. In particular, for the
$4^1\Sigma^+_{\mathrm u}$ state under consideration the
theoretical calculations predict a double minimum potential, with
a deep inner well and a shallow but broad outer one, in addition
to a clearly visible inflection in the right-hand wall of the
inner well. Our experimental data related to the
$4^1\Sigma^+_{\mathrm u}$ state contained 277 energies of levels
located in the inner well; they allowed us to find the detailed
shape of this well. On the other hand, only 55 levels have been
observed in the region above the potential barrier, whereas the
outer potential well was totally inaccessible in the present
experiment. Therefore determination of the outer well turned out
to be an ill-posed problem. We have shown, however, that a
physically sensible solution can be found by application of a
simple but effective regularization procedure, based on an
additional constraint imposed on a constructed potential. As a
result, a smooth potential energy curve corresponding to a broad
range of internuclear distances (2.4 to 20.0 \AA) has been
generated, reproducing energies of all experimental rovibronic
levels to within the experimental accuracy and displaying the
expected physical behavior.

In Section II we present the experimental setup and techniques
used. Section III contains a brief description of the recorded
spectra. The basic concepts of the inverted perturbation approach
as well as the proposed modification of the algorithm are
discussed in Section IV and their application to deduction of the
potential energy curve for the 4$^1\Sigma^+_{\mathrm u}$ state is
shown in Section V. Finally, Section VI summarizes our results.

\section{Experimental}

To study the  $4^1\Sigma^+_{\mathrm u} \leftarrow $
X$^1\Sigma^+_{\mathrm g}$ system in Na$_2$, we employed the
V-type optical-optical double resonance polarization labeling
spectroscopy technique with two independent pump and probe light
beams. In our version of the method the probe beam had a fixed
frequency and excited a few assigned molecular transitions,
whereas the pump beam was tuned over an investigated spectrum.
The experimental apparatus and method have been described in
detail elsewhere \cite{a1,a2} and therefore only the essential
features will be presented here. Na$_2$ molecules were produced
in a heatpipe oven, operating at around 750 K, with 4 Torr of
helium as a buffer gas. The pump and probe light beams were
superimposed collinearly in the sodium vapor zone. As a probe
light we employed one of the five blue-green lines of a linearly
polarized multimode Ar$^+$ laser (Carl Zeiss ILM 120, $\lambda =$
476.5, 488.0, 496.5, 501.7 and 514.5 nm), with a typical power
ranging from 20 to 150 mW. The probe laser light excited several
known transitions in the B$^1\Pi_{\mathrm u} \leftarrow$
X$^1\Sigma^+_{\mathrm g}$ system of sodium dimer \cite{c}, thus
labeling the involved rovibrational levels in the ground
electronic state. A parametric oscillator/amplifier system
(Sunlite EX, Continuum) provided with a frequency doubler (FX-1)
and pumped with the third harmonic of an injection seeded Nd:YAG
laser (Powerlite 8000) served as a source of the pump light. The
system produced UV radiation with a typical energy 3 mJ per pulse
and a spectral width of 0.16 \rcm. The tunable pump light excited
sodium molecules from the ground X$^1\Sigma^+_{\mathrm g}$ state
to the $4^1\Sigma^+_{\mathrm u}$ state studied in this
experiment. The frequency of the pump beam was calibrated against
the optogalvanic spectrum of argon in a hollow cathode discharge
tube and the transmission fringes of a Fabry-P\'{e}rot
interferometer 0.5 cm long. The uncertainty in determining the
line centers for strong molecular lines is estimated to 0.05 \rcm.

Crossed polarizers were placed at
both sides of the heatpipe oven in the path of the probe beam. At
the frequencies at which transitions induced by the pump beam
shared the same lower levels with the probe transitions, the
probe light passed through the analyzer. The signal was recorded
with a photomultiplier and processed with a computer. For
measurements of polarization spectra labeled via P or R
transitions, the pump beam was circularly polarized; for
polarization spectra labeled via Q transitions, the linear
polarization of the pump beam was chosen \cite{b}.

 \begin{figure}
  \centering
\epsfig{file=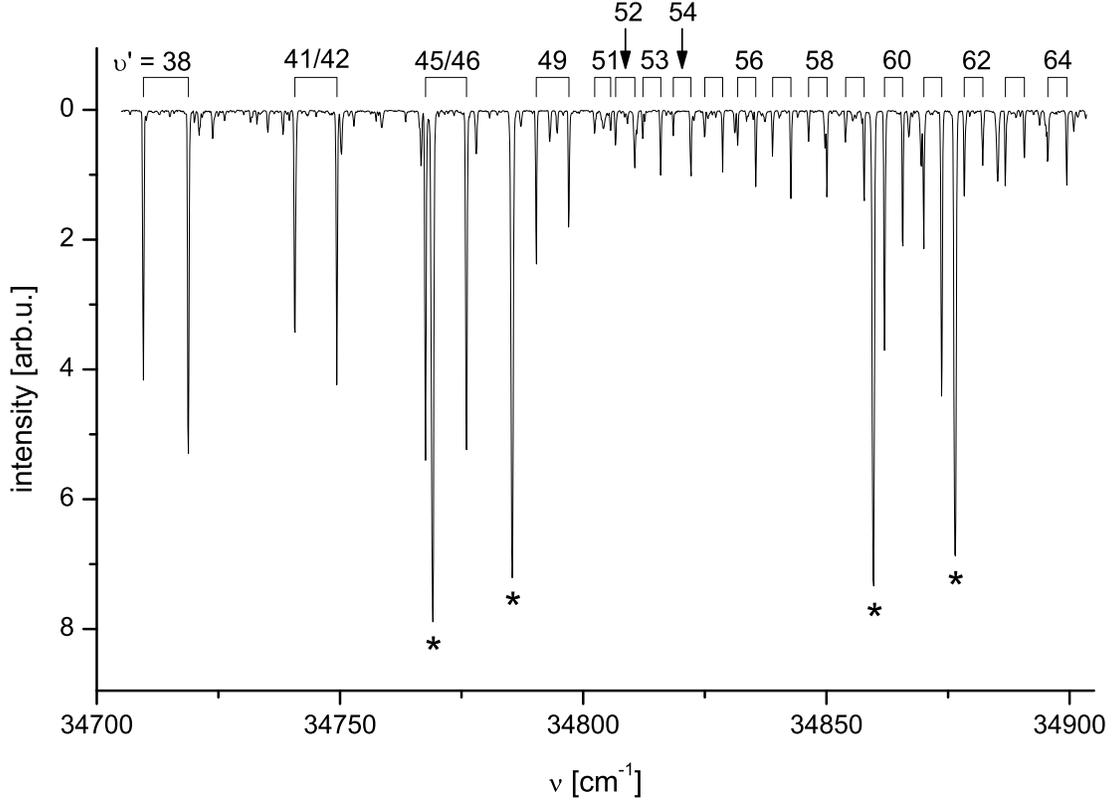,width=\linewidth}
 \caption{Part of the polarization spectrum of Na$_2$ obtained
 with the 501.7 nm line of the Ar$^+$ laser as the probe and
circularly polarized pump light. The assigned $v'$ progressions
correspond to transitions $4^1\Sigma^+_{\mathrm u}$($v'$, $J' =
J'' \pm$ 1) $\leftarrow $ X$^1\Sigma^+_{\mathrm g}$($v'' = 2, J''
=$ 43). Note that because of near resonance of vibrational levels
in the inner and outer potential wells, in two observed doublets
$v'$ numbers differ by one for P and R lines. A systematic
disparity of intensities of P and R lines for $v' = 52$ to 59 is
presently not understood. Lines denoted by stars correspond to
D$^1\Pi_{\mathrm u}\leftarrow $ X$^1\Sigma^+_{\mathrm g}$
transition in Na$_2$, overlapping the investigated band system.}
  \label{Fig1}
   \end{figure}

\section{Results}

\begin{figure}
  \centering
\epsfig{file=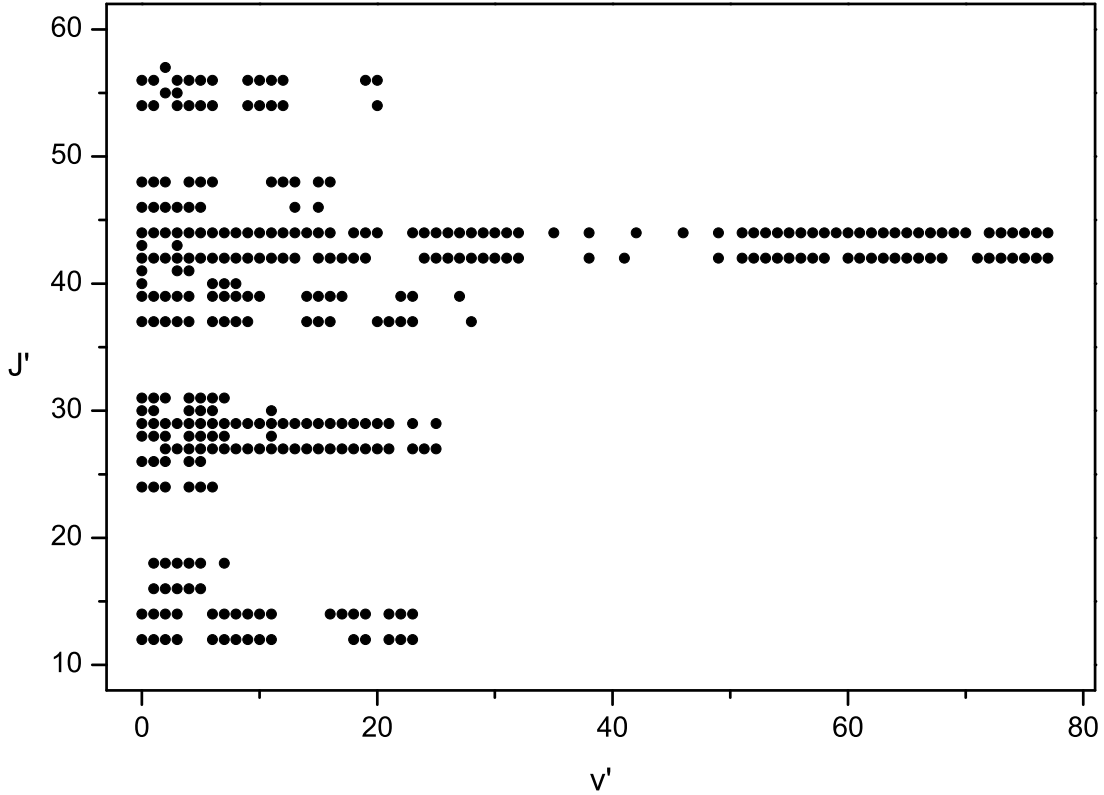,width=\linewidth}
 \caption{Range of rovibrational levels used in the analysis of the
 $4^1\Sigma^+_{\mathrm u}$ state in Na$_2$.}
  \label{Fig2}
   \end{figure}

We recorded the polarization spectrum of the
$4^1\Sigma^+_{\mathrm u} \leftarrow$ X$^1\Sigma^+_{\mathrm g}$
system of Na$_2$ in a range between 32100 and 35100 \rcm.
Fig.~\ref{Fig1} shows a particularly interesting fragment of the
spectrum where an abrupt change of vibrational spacing in the
observed $v'$ progression corresponds to transition from the
inner well to a region above the potential barrier. The analyzed
spectral lines provided information about 332 rovibrational
levels in the 4$^1\Sigma^+_{\mathrm u}$  state. The data field is
illustrated in Fig.~\ref{Fig2}. The highest vibrational level
identified by us corresponds to $v' = 77$ (\textit{vide infra})
whereas the rotational quantum numbers $J'$ are spread between 12
and 56. Assignment of $v' = 0$ level was based on an assumption
that this was the lowest level observed. After Franck-Condon
factors for transitions from the ground X$^1\Sigma^+_{\mathrm g}$
state were calculated, comparison of them with the measured
relative strengths of the spectral lines confirmed the
assignment. It must be noted that the outer well in
4$^1\Sigma^+_{\mathrm u}$ was inaccessible in the present
experiment because of negligible overlap between vibrational wave
functions located there and those corresponding to the ground
state levels. The measured wave numbers of lines have been
converted to energies of 4$^1\Sigma^+_{\mathrm u}$ state levels
referred to the bottom of the X$^1\Sigma^+_{\mathrm g}$ state
potential well, using the ground state molecular constants of
Kusch and Hessel \cite{c}. As they reproduce energies of the
rovibrational levels in the ground state with an accuracy
exceeding the precision of our measurements, no additional errors
were introduced into our analysis of the 4$^1\Sigma^+_{\mathrm
u}$ state.

\section{The IPA algorithm using a pointwise potential and its modification}

The main idea of the IPA is to start with an approximate potential energy
curve $U_0(R,\bm{a})$ with eigenenergies $E^0_i$ and eigenfunctions $\Psi_i^0$.
The fitting parameters $\bm{a}=(a_1,a_2,...,a_M)$ are used in order to modify $U_0(R,\bm{a})$ in
such a way that its new eigenenergies $E_i$
agree with the experimental  ones $E^{\mathrm{exp}}_i$ in the LSA
sense \cite{Kozman}. If the approximate potential is close to the ``true'' one, we can
assume a linear dependence of the eigenenergies on small changes of the
parameters $a_j$

  \begin{equation}
\label{1}
       E_i = E^0_i + \sum_j \frac{\partial E_i}{\partial a_j} \Delta
       a_j \mbox{ ;}
   \end{equation}

\noindent $E^0_i$ and $\frac{\partial E_i}{\partial a_j}$ denote
the eigenvalue and its first derivatives with respect to $a_j$
for the approximate potential, i.e. at $\bm{a}=\bm{a}^0$.
Application of the Hellman-Feyman theorem leads to

 \begin{equation}
        \frac{\partial E_i}{\partial a_j} =
        <\Psi^0_i | \frac{\partial U(R,\bm{a})}{\partial a_j} | \Psi^0_i >.
 \end{equation}

In order to parametrize the molecular potential we use an
equidistant grid of points $(R_k,u_k)$ connected with a cubic
spline function \cite{weCPC}. Values of the potential at $R=R_k$,
that is $u_k$, are fitting parameters. In this case it is possible
to show \cite{weCPC} that

\begin{equation}
\label{2}
           U(R,\bm{u})=\sum_k S_k(R)u_k \mbox{ ,}
\end{equation}

\noindent where $S_k(R)$ are known functions of $R$. Since

\begin{equation}
         \frac{\partial U(R,\bm{u})}{\partial u_j} = S_j(R) \mbox{ ,}
\end{equation}

\noindent Eq.~\ref{1} can be rewritten as

\begin{equation}
\label{3}
         E_i = E^0_i + \sum_j <\Psi^0_i | S_j(R) | \Psi^0_i > \Delta u_j.
\end{equation}

\noindent where $\Delta u_j$ is the correction to the $j$-th
fitting parameter. Replacing $E_i$ with $E^{\mathrm{exp}}_i$
provides a set of $N$ linear equations ($N$ denotes the number of
experimental levels) with $M$ unknowns (where $M$ is the number
of the fitted parameters), which should be solved in the LSA
sense.

Usually the whole spectrum of eigenvalues is not available and
therefore some parts of the potential cannot be reliably
retrieved from the experimental data. In case of the pointwise
representation of the potential this means that the calculated
term energies of the experimentally observed levels are not
sensitive to some of the $u_k$. The inversion problem becomes
ill-conditioned then and its solution may be unstable to data
perturbation and lacking a physical sense, particularly in terms
of a desired smoothness of the constructed potential. The
ill-conditioning of the problem does not imply that a meaningful
approximate solution cannot be found, but an extra care has to be
taken to obtain a physically acceptable result. The singular
value decomposition (SVD) method \cite{SVD} chosen by us for
solving the initial system of linear equations offers a partial
remedy. Also a sensible choice of a sparse grid for the pointwise
potential in the regions badly characterized by the experimental
data can flatten to some extent the undesirable wiggles on it
(see \cite{we6S}). Here we propose a simpler, more general and
more effective solution, based on an additional constraint
imposed on the constructed potential.

Generally, the usual way of obtaining a smooth fitted function
$f(R,\bm{a})$ is to add to the merit function $\chi^2(\bm{a})$  a
regularizing functional $H(\bm{a})$ which is responsible for the
smoothness of the solution and to minimize the sum

\begin{equation}
\mathrm{minimize}\mbox{: }  \chi^2(\bm{a}) + \lambda^2 H(\bm{a})
\end{equation}

\noindent where $\lambda$ is a parameter used to tune the degree
of regularization. A possible form of $H(\bm{a})$ is (see e.g.
\cite{SVD}, Chapter 18):

\begin{equation}
\label{4} H(\bm{a})=\int_{R_1}^{R_2} (f''(R, \bm{a}))^2 dR
\end{equation}

Here, by minimizing the integral of the square of the second derivative,
one sets the additional condition that the fitted function should be as
close to a straight line as possible between $R_1$ and $R_2$.
Since in our version of the
IPA method the fitted potential is defined as a cubic spline
function drawn between given grid points $(R_k, u_k)$, the second
derivative $U''(R, \bm{u})$ between these points is a linear function
of $R$. Therefore we define a somewhat simplified form of the
regularizing functional as

\begin{equation}
H(\bm{u}) = \sum_j (U''(R_j,\bm{u}))^2
\end{equation}

\noindent where $U''(R_j,\bm{u})= U''_j$ denotes the second derivative
of the potential in a grid point $R_j$.

Similarly to Eq.~(\ref{2}), in Ref.~\cite{weCPC} it was shown that

\begin{equation}
\label{9} U''_j=\sum_i L_{ji} u_i.
\end{equation}

\noindent where $L_{ji}$ are known coefficients.
Therefore in the
present case the regularization condition is reduced to supplementing
the system of linear equations (\ref{3}) by a set of M equations:

\begin{equation}
\label{reg} \sum_i \lambda L_{ji} (u^0_i+ \Delta u_i)=0
\end{equation}

Here $u_i^0$ is the initial value of the $i$-th fitting parameter.
Each equation requires that the respective second derivative
$U''_j$ equals zero and the summation is performed over all fitted
parameters. When $\lambda=0$, the system (\ref{3}) remains
unchanged, i.e. no regularization is imposed.  Increasing
$\lambda$ results in flattening of the potential in competition
with the initial condition defined by (\ref{3}). In principle the
value of $\lambda$ in (\ref{reg}) can depend on $j$, varying
smoothly from zero for the parts of the potential with abundant
experimental data to some large value for the regions badly
characterized by the experiment; the other sensible possibility is
to make the change stepwise and this approach has been adopted in
the present work.

\section{Potential curve of the 4$^1\Sigma^{+}_{\mathrm u}$ state}

With the body of data described in Section III we applied the IPA
technique to determine the potential energy curve of the
4$^1\Sigma^{+}_{\mathrm u}$ state. Initially we limited our
analysis to the inner well of the molecular potential, containing
a majority of levels observed in the present experiment.
Transitions to rovibrational levels located there formed clear
vibrational progressions of spectral lines. However, we observed
that distances between subsequent P, R doublets did not decrease
in a systematic way even for the first few doublets in each
progression, as expected for a well-behaved molecular potential.
This observation supported theoretical prediction of substantial
deformation of the inner well. In consequence, levels situated in
the inner well could not be characterized by a set of molecular
constants and a standard RKR procedure could not be used for
determination of the inner well. Under these circumstances we
rather applied a variant of the RKR method first proposed by
Stwalley \cite{Stwalley}, which allows to construct an approximate
potential energy curve from vibrational term values $G$($v$) and
rotational constants $B_v$ obtained directly from the
experimental spectrum, i.e. from a distance of the subsequent P, R
doublets and the P-R combination differences measured in a chosen
progression, preferably being the longest one.

The potential curve generated in this way was used then as an
approximate starting potential for the standard IPA procedure
\cite{weCPC} (i.e. without regularization). It provided a refined
potential energy curve which reproduced eigenenergies of levels
distant from the top of the potential barrier by more than 50
\rcm\ (i.e. the levels, for which the presence of the outer well
could be neglected) within the experimental accuracy. In the
second part of our analysis we used a larger data set including
levels close to and above the potential barrier. In this case the
existence of both potential wells has to be taken into account
and accordingly we extended the region of internuclear distances
$R$, for which the potential was considered, to $R \approx 16$
\AA. As a starting potential for the next run of the IPA routine
we adopted a hybrid potential consisting of the inner well from
the previous step matched smoothly with the theoretical curve
calculated by Magnier \cite{e} used to represent the outer well.

An equidistant grid was used for the approximate potential since
Eqs.~\ref{2} and \ref{9} are derived under this condition.
Although a generalization of both equations is possible also for a
non-equidistant grid, this would require significant changes in
the fitting code \cite{weCPC}. Moreover, it is not clear
\textit{a priori} how to find the optimal distribution of the grid
points. Hence initially we defined $U_0(R)$ in 110 equidistant
points for 2.4 \AA $ \leq R \leq $ 15.9 \AA, i.e. a grid dense
enough to describe the steepest changes of the potential curve.

After several iterations the r.m.s. deviation of the fit
decreased below 0.05 \rcm, that is the generated IPA potential
reproduced all the observed levels with an expected accuracy. At
closer inspection, however, the potential turned out
unsatisfactory: the outer well displayed unphysical
irregularities in a form of ripples, resulting from scarcity of
experimental data defining the region of large $R$ values.
Therefore we refitted the potential including the regularization
condition in the form of Eq.~\ref{reg}. During the fit $\lambda$
was varied between $1.0$ and $4.0$ for $R > R_{\mathrm{reg}}$ and
set to zero otherwise. Several values of $R_{\mathrm{reg}}$
between $6.5$ and $8.0$ \AA\ were tested. The value of $\lambda$
was gradually increased to smooth the outer potential well, until
the quality of the fit started to deteriorate. Once a
satisfactory shape of the outer well was achieved, we reduced the
number of grid points as described in Ref.~\cite{weCPC}.

 \begin{figure}
 \centering
\epsfig{file=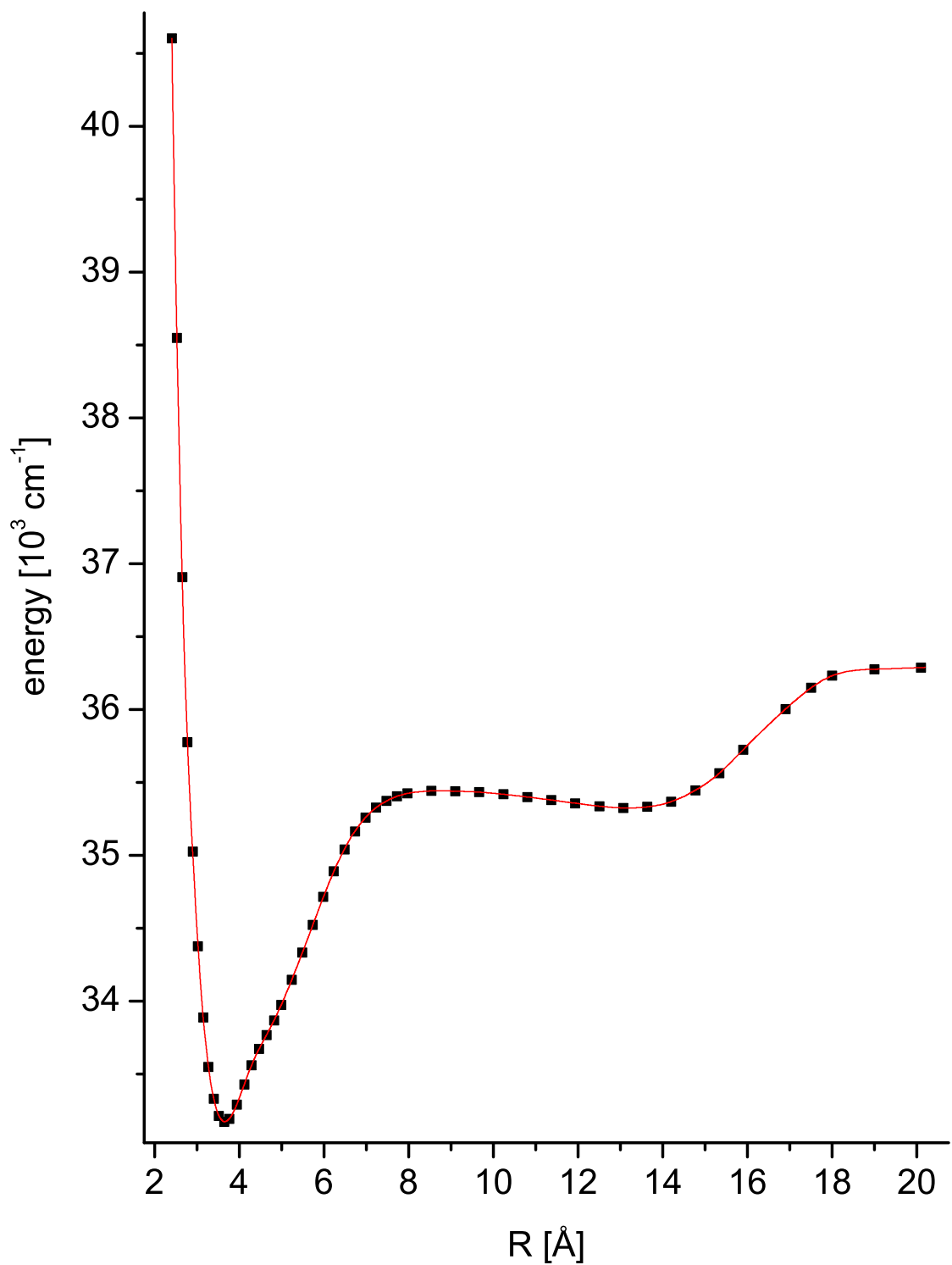,width=\linewidth}
 \caption{The generated potential curve of the $4^1\Sigma^+_{\mathrm u}$
 state. Zero on the energy scale corresponds to the bottom of the
 ground X$^1\Sigma^+_{\mathrm g}$ potential \cite{c}.}
  \label{Fig3}
   \end{figure}

The final version of the potential is presented in Table I and in
Fig.~\ref{Fig3}. The last five points were taken from theoretical
calculations \cite{e}. Their influence on the energy level
positions is negligible and they were added mainly to ensure
proper boundary conditions for solving the \Sch\ equation. In
order to calculate the value of the potential for arbitrary $R$,
a natural cubic spline should be used \cite{SVD}. The potential
reproduces 325 of totally 332 observed energy levels with a
standard deviation of 0.044 \rcm\ and a normalized standard
deviation of 0.88. Seven levels were excluded from the fit since
the deviations from the calculated term energies exceeded 0.15
\rcm. We attribute these discrepancies to perturbations by the
neighboring singlet or triplet states \cite{e}.

\begin{table*}
\label{ipapot}
\caption{The IPA potential energy curve of the $4^1\Sigma^+_{\mathrm u}$ state
          in Na$_2$.}
 \begin{tabular*}{0.9\linewidth}{@{\extracolsep{\fill}}rrrr}\hline
 R [\AA] & U [cm$^{-1}$]& R [\AA] & U [cm$^{-1}$]\\ \hline
  2.4000 & 40604.123 &      6.7257  &    35166.484\\
  2.5239 & 38549.528 &      6.9736  &    35258.930\\
  2.6477 & 36908.022 &      7.2216  &    35327.723\\
  2.7716 & 35776.823 &      7.4695  &    35373.419\\
  2.8954 & 35026.062 &      7.7175  &    35405.959\\
  3.0193 & 34379.006 &      7.9650  &    35425.336\\
  3.1431 & 33888.597 &      8.5318  &    35442.782\\
  3.2670 & 33549.507 &      9.0986  &    35441.533  \\
  3.3908 & 33332.544 &      9.6654   &   35434.522  \\
  3.5147 & 33214.236 &     10.2321   &  35419.888   \\
  3.6385 & 33174.430 &     10.7989   &  35401.202   \\
  3.7624 & 33195.060 &     11.3657   &  35380.528   \\
  3.9359 & 33291.770 &     11.9325   &  35358.744   \\
  4.1119 & 33428.812 &     12.4993   &  35338.612   \\
  4.2878 & 33562.727 &     13.0661   &  35326.650   \\
  4.4638 & 33673.469 &     13.6329   &  35333.401    \\
  4.6397 & 33769.607  &    14.1996   &  35369.969     \\
  4.8157 & 33868.489  &    14.7664   &  35446.576     \\
  4.9900 & 33976.789  &    15.3332   &  35564.687     \\
  5.2380 & 34148.711  &    15.9000   &  35726.822     \\
  5.4859 & 34333.616  &    16.9000   &  36005.000     \\
  5.7339 & 34525.107  &    17.5000   &  36150.000     \\
  5.9818 & 34715.356  &    18.0000   &  36233.000     \\
  6.2298 & 34890.866  &    19.0000   &  36277.700     \\
  6.4777 & 35042.921  &    20.0000   &  36284.700  \\
\hline
\end{tabular*}

\end{table*}

\section{Final remarks}

In the present work we attempted to construct the highly irregular
potential energy curve of the 4$^1\Sigma^{+}_{\mathrm u}$ state
in Na$_2$ from the experimental data abundant for levels in the
inner potential well but scarce for the region around and above
the internal barrier. A version of the IPA procedure involving the
pointwise representation of the potential and its regularization
was used. The inner well has been determined accurately, however
problems arose when trying to find the shape of the barrier and
the outer potential well. A small number of the measured level
energies influencing these parts of the potential suggested that
determination of its shape was not possible. In mathematical
terms, inversion of the spectroscopic data to a potential curve
could not provide a unique result. However, from many possible
potentials, representing positions of all the experimental levels
equally well, we were able to select one with a sensible physical
shape. A general procedure, based on the mathematical concept of
regularization of the constructed potential curve, has been
suggested to deal with such problems. Still, it should be borne
in mind that the outer well of the potential reported in Table I
is only a plausible smooth solution to the inversion problem and
it would be unreasonable to speculate about its accuracy. The
only way to improve its reliability is to collect more
experimental data related to this part of the potential.

It should be also noted that the uncertainty of the outer
potential well affects the vibrational numbering of the
rovibrational levels of the 4$^1\Sigma^{+}_{\mathrm u}$ state.
Small changes of the outer well may cause a change of the
numbering of levels which wave functions are significantly
nonzero mainly in the inner well. Of course this will change
neither their term energies, nor the relative vibrational
numbering of the levels belonging to the inner well, i.e. the
quantities directly observable in the present experiment.

\begin{acknowledgments}
This work has been funded in part by grant No.~2~P03B~063~23 from
the Polish Committee for Scientific Research. A.P. acknowledges a
support from the Center of Excellence ASPECT (program
"Competitive and Sustainable Growth", G6MA-CT-2002-04021).
\end{acknowledgments}

\end{document}